# What is a Question?

Kevin H. Knuth

*NASA Ames Research Center, Computational Sciences Department, Code IC*
*Moffett Field CA 94035 USA*

**Abstract.** A given question can be defined in terms of the set of statements or assertions that answer it. Application of logical inference to these sets of assertions allows one to derive the logic of inquiry among questions. There are interesting symmetries between the logics of inference and inquiry; where probability describes the degree to which a premise implies an assertion, there exists an analogous measure that describes the bearing or relevance that a question has on an outstanding issue. These have been extended to suggest that the logic of inquiry results in functional relationships analogous to, although more general than, those found in information theory.

Employing lattice theory, I examine in greater detail the structure of the space of assertions and questions demonstrating that the symmetries between the logical relations in each of the spaces derive directly from the lattice structure. Furthermore, I show that while symmetries between the spaces exist, the two lattices are not isomorphic. The lattice of assertions is described by a Boolean lattice $\mathbf{2}^N$, whereas the lattice of assuredly real questions is shown to be a sublattice of the free distributive lattice $\mathbf{FD}(N) = \mathbf{2}^{\mathbf{2}^N}$. Thus there does not exist a one-to-one mapping of assertions to questions, there is no reflection symmetry between the two spaces, and questions in general do not possess complements. Last, with these lattice structures in mind, I discuss the relationship between probability, relevance, and entropy.

> "*Man has made some machines that can answer questions provided the facts are profusely stored in them, but we will never be able to make a machine that will ask questions. The ability to ask the right question is more than half the battle of finding the answer.*"
> - Thomas J. Watson (1874-1956)

## INTRODUCTION

It was demonstrated by Richard T. Cox [1, 2] that probability theory represents a generalization of Boolean implication to a degree of implication represented by a real number. This insight has placed probability theory on solid ground as a calculus for conducting inductive inference. While at this stage this work is undoubtedly his greatest contribution, his ultimate paper, which takes steps to derive the logic of questions in terms of the set of assertions that answer them, may prove yet to be the most revolutionary. While much work has been done extending and applying Cox's results [3-12], the mathematical structure of the space of questions remains poorly understood. In this paper I employ lattice theory to describe the structure of the space of assertions and demonstrate how logical implication on the Boolean lattice provides the framework on which the calculus of inductive inference is constructed. I then introduce questions by following Cox [13] who defined a question in terms of the set of assertions that can answer it. The lattice structure of questions is then explored and the calculus for manipulating the relevance of a question to an unresolved issue is examined.

The first section is devoted to the formalism behind the concepts of partially ordered sets and lattices. The second section deals with the logic of assertions and introduces Boolean lattices. In the third section, I introduce the definition of a question and introduce the concept of an ideal question. From the set of ideal questions I construct the entire question lattice identifying it as a free distributive lattice. Assuredly real questions are then shown to comprise a sublattice of the entire lattice of questions. In the last section I discuss the relationship between probability, relevance, and entropy in the context of the lattice structure of these spaces.

# FORMALISM

## Partially Ordered Sets

In this section I begin with the concept of a partially ordered set, called a *poset*, which is defined as a set with a binary ordering relation denoted by $a \leq b$, which satisfies for all $a$, $b$, $c$ [14]:

P1.     For all $a$, $a \leq a$.                                                       (Reflexive)
P2.     If $a \leq b$ and $b \leq a$, then $a = b$                             (Antisymmetry)
P3.     If $a \leq b$ and $b \leq c$, then $a \leq c$                             (Transitivity)

Alternatively one can write $a \leq b$ as $b \geq a$ and read "$b$ contains $a$" or "$b$ includes $a$". If $a \leq b$ and $a \neq b$ one can write $a < b$ and read "$a$ is less than $b$" or "$a$ is properly contained in $b$". Furthermore, if $a < b$, but $a < x < b$ is not true for any $x$ in the poset $P$, then we say that "$b$ covers $a$", written $a \prec b$. In this case $b$ can be considered an immediate superior to $a$ in a hierarchy. The set of natural numbers {1, 2, 3, 4, 5} along with the binary relation "less than or equal to" $\leq$ is an example of a poset. In this poset, the number 3 covers the number 2 as $2 < 3$, but there is no number $x$ in the set where $2 < x < 3$. This covering relation is useful in constructing diagrams to visualize the structure imposed on these sets by the binary relation.

To demonstrate the construction of these diagrams, consider the poset defined by the powerset of $\{a,b,c\}$ with the binary relation $\subseteq$ read "is a subset of", $P = (\{\varnothing, \{a\}, \{b\}, \{c\}, \{a,b\}, \{b,c\}, \{a,c\}, \{a,b,c\}\}, \subseteq)$ where the powerset $\wp(X)$ of a set $X$ is the set of all possible subsets of $X$. As an example, it is true that $\{a\} \subseteq \{a,b,c\}$, read "$\{a\}$ is included in $\{a,b,c\}$". Furthermore, it is true that $\{a\} \subset \{a,b,c\}$, read "$\{a\}$ is properly contained in $\{a,b,c\}$" as $\{a\} \subseteq \{a,b,c\}$, but

$\{a\} \neq \{a,b,c\}$. However, $\{a,b,c\}$ does not cover $\{a\}$ as $\{a\} \subset \{a,b\} \subset \{a,b,c\}$. We can construct a diagram (Figure 1) by choosing two elements *x* and *y* from the set, and writing *y* above *x* when $x \subset y$. In addition, we connect two elements *x* and *y* with a line when *y* covers *x*, $x \prec y$.

Posets also possess a duality in the sense that the converse of any partial ordering is itself a partial ordering [14]. This is known as the *duality principle* and can be understood by changing the ordering relation "is included in" to "includes" which equates graphically to flipping the poset diagram upside-down.

With these examples of posets in mind, I must briefly describe a few more concepts. If one considers a subset *X* of a poset *P*, we can talk about an element $a \in P$ that contains every element $x \in X$; such an element is called an *upper bound* of the subset *X*. The *least upper bound*, or l.u.b., is an element in *P*, which is an upper bound of *X* and is contained in every other upper bound of *X*. Thus the l.u.b. can be thought of as the immediate successor to the subset *X* as one moves up the hierarchy. Dually we can define the *greatest lower bound*, or g.l.b. The *least element* of a subset *X* is an element $a \in X$ such that $a \leq x$ for all $x \in X$. The *greatest element* is defined dually.

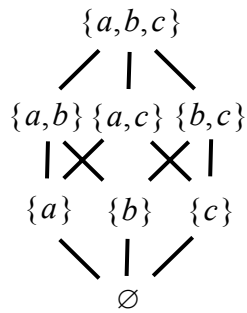

**FIGURE 1.** The poset $P = (\{\varnothing, \{a\}, \{b\}, \{c\}, \{a,b\}, \{b,c\}, \{a,c\}, \{a,b,c\}\}, \subseteq)$ results in the diagram shown here. The binary relation $\subseteq$ dictates the height of an element in the diagram. The concept of covering allows us to draw lines between a pair of elements signifying that the higher element in the pair is an immediate successor in the hierarchy. Note that $\{a\}$ is covered by two elements. These diagrams nicely illustrate the structural properties of the poset. The element $\{a,b,c\}$ is the greatest element of *P* and $\varnothing$ is the least element of *P*.

## Lattices

The next important concept is the *lattice*. A lattice is a poset *P* where every pair of elements *x* and *y* has a least upper bound called the *join*, denoted as $x \vee y$, and a greatest lower bound called the *meet*, denoted by $x \wedge y$. The meet and join obey the following relations [14]:

L1.     $x \wedge x = x, \quad x \vee x = x$                                                    (Idempotent)

L2.     $x \wedge y = y \wedge x, \quad x \vee y = y \vee x$                                 (Commutative)

L3.     $x \wedge (y \wedge z) = (x \wedge y) \wedge z, \quad x \vee (y \vee z) = (x \vee y) \vee z$       (Associative)

L4.     $x \wedge (x \vee y) = x \vee (x \wedge y) = x$                                (Absorption)

In addition, for elements $x$ and $y$ that satisfy $x \leq y$ their meet and join satisfy *the consistency relations*

C1.    $x \wedge y = x$           ($x$ is the greatest lower bound of $x$ and $y$)
C2.    $x \vee y = y$            ($y$ is the least upper bound of $x$ and $y$).

The relations L1-4 above come in pairs related by the duality principle; as they hold equally for a lattice $L$ and its dual lattice (denoted $L^\partial$), which is obtained by reversing the ordering relation thus exchanging upper bounds for lower bounds and hence exchanging joins and meets. Note that the meet and join are generally defined for all posets satisfying the definition of a lattice; even though the notation is the same they should not be confused with the logical conjunction and disjunction, which refer to a specific ordering relation. I will get to how they are related and we will see that lattice theory provides a general framework that clears up some mysteries surrounding the space of assertions and the space of questions.

# THE LOGIC OF ASSERTIONS

## Boolean Lattices

I introduce the concept of a Boolean lattice, which possesses structure in addition to L1-4. A Boolean lattice is a *distributive lattice* satisfying the following identities for all $x, y, z$:

B1.    $x \wedge (y \vee z) = (x \wedge y) \vee (x \wedge z)$           (Distributive)
       $x \vee (y \wedge z) = (x \vee y) \wedge (x \vee z)$

Again the two identities are related by the duality principle. Last the Boolean lattice is a *complemented lattice*, such that each element $x$ has one and only one *complement* $\sim x$ that satisfies [14]:

B2.    $x \wedge \sim x = O$       $x \vee \sim x = I$
B3.    $\sim (\sim x) = x$
B4.    $\sim (x \wedge y) = \sim x \vee \sim y$       $\sim (x \vee y) = \sim x \wedge \sim y$

where $O$ and $I$ are the least and greatest elements, respectively, of the lattice. Thus a Boolean lattice is a *complemented distributive lattice*.

We now consider a specific application where the elements $a$ and $b$ are logical assertions and the ordering relation is $x \leq y \equiv x \rightarrow y$, read "$x$ implies $y$". The logical operations of conjunction and disjunction can be used to generate a set of four logical statements, which with the binary relation "implies" forms a Boolean lattice displayed in Figure 2. It can be shown that the meet of $a$ and $b$, written $a \wedge b$, is identified with the logical conjunction of *a* and *b*, and the join of *a* and *b*, written $a \vee b$, is identified with the logical disjunction of *a* and *b*. I will require that the lattice be complemented, which means that the complement of *a* must be *b*, $\sim a = b$, and vice versa. If we require the assertions to be exhaustive, then either *a* or *b* are true, and their join, the disjunction $a \vee b$, must always be true. By B2 $a \vee b$ must be the greatest element and is thus *I*, which in logic is called *the truism*, as it is always true. Similarly their meet, the conjunction $a \wedge b$, is the least element *O* and when *a* and *b* are mutually exclusive *O* must always be false, earning it the name *the absurdity*.

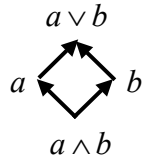

**FIGURE 2**. The lattice diagram formed from two assertions *a* and *b*. In this diagram I chose to use arrows to emphasize the direction of implication among the assertions in the lattice.

The symbol for the truism *I* mirrors the *I* used by Jaynes to symbolize "one's prior information" [15]. In fact, in an inference problem, if one believes that one of a set of assertions is true then one's prior knowledge consists, in part, of the fact that the disjunction of the entire set of assertions is true. By fortuitous circumstance the notation of lattice theory agrees quite nicely with the notation used by Jaynes.

Deductive inference refers to the process where one knows that an assertion *a* is true, and deduces that any assertion reached by a chain of arrows must also be true. If for two assertions *x* and *y* elements of a lattice *L*, *x* is included in *y*, $x \leq y$, we say that *x* implies *y*, denoted $x \rightarrow y$.

If a set of assertions used to generate the lattice is a mutually exclusive set then all possible conjunctions of these assertions are equal to the absurdity,

$$x \wedge y = O \quad \text{for all } x, y \in \ : x \neq y.$$

These elements that cover *O* are called *atoms* or *points*. As all other elements are formed from joins of these atoms, they are called generators or generating elements and the lattice is called an *atomic lattice*. The total number of assertions in the atomic Boolean lattice is $2^N$, where *N* is the number of atoms. These Boolean lattices can be named according to the number of atoms, $\mathbf{2}^N$. The first three atomic Boolean lattices are shown in Figure 3. In these figures one can visualize the curious fact of logic: the absurdity *O* implies everything. Also, it is instructive to identify and verify the complements of the generators (eg. in $\mathbf{2}^2$, $\sim a = b$, and in $\mathbf{2}^3$, $\sim a = b \vee c$). These lattices are self-dual as the same lattice structure results by reversing the ordering relation (turning the diagram upside-down) and interchanging meets and joins ($x \vee y$ and $x \wedge y$).

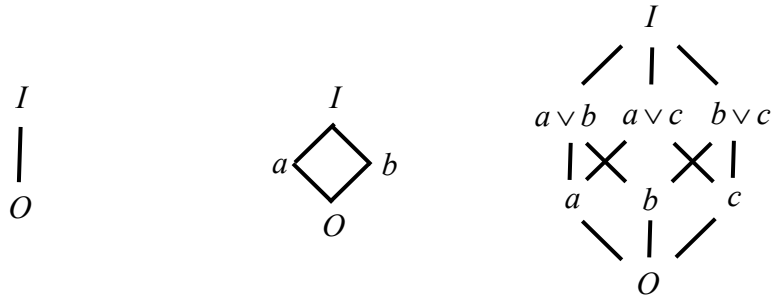

**FIGURE 3**. Here are the first three atomic Boolean lattices where the upward pointing arrows denoting the property "*is included in*" or "implies" have been omitted. Left: The lattice $\mathbf{2}^1$ where $I = a$. Center: The lattice $\mathbf{2}^2$ generated from two assertions (same as Fig. 2) where $O = a \wedge b$ and $I = a \vee b$. Right: The lattice $\mathbf{2}^3$ generated from three atomic assertions where the conjunction of all three assertions is represented by the absurdity $O$, and the disjunction of all three assertions is represented by the truism $I$.

For fun we could consider creating another lattice $\Lambda^N$ where we define each atom $\lambda_i$ in $\Lambda^N$ from the mapping $\Lambda : b_i \rightarrow \lambda_i = \{b_i\}$ as a set containing a single atomic assertion $b_i$ from $\mathbf{2}^N$. In addition, we map the operations of logical conjunction and disjunction to set intersection and union respectively, that is $(\mathbf{2}^3, \wedge, \vee) \rightarrow (\Lambda^3, \cap, \cup)$. Figure 4 shows $\Lambda^3$ generated from $\mathbf{2}^3$. As we can define a one-to-one and onto mapping (an *isomorphism*) from $\mathbf{2}^3$ to $\Lambda^3$, the lattices $\Lambda^3$ and $\mathbf{2}^3$ are said to be *isomorphic*, which I shall write as $\Lambda^3 = \mathbf{2}^3$. The Boolean nature of the lattice $\Lambda^3$ can be related to a base-2 number system by visualizing each element in the lattice as being labeled with a set of three numbers, each either a one or zero, denoting whether the set contains (1) or does not contain (0) each of the three atoms. $\{a,b,c\}$

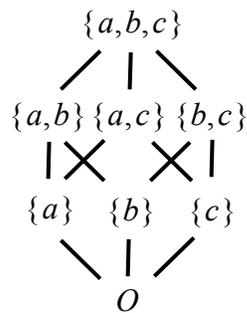

**FIGURE 4**. The lattice $\Lambda^3$ was generated from $\mathbf{2}^3$ by defining each atom as a set containing a single atomic assertion from $\mathbf{2}^3$, and by replacing the operations of logical conjunction and disjunction with set intersection and union, respectively as in $(\mathbf{2}^3, \wedge, \vee) \rightarrow (\Lambda^3, \cap, \cup)$. Note that in this lattice $I = \{a,b,c\}$ and $O = \emptyset$ (the empty set). As there is a one-to-one and onto mapping of this lattice to the lattice in Fig. 3 (right), they are isomorphic.

# Inductive Inference guided by Lattices

Inductive inference derives from deductive inference as a generalization of Boolean implication to a relative degree of implication. In the lattice formalism that this is equivalent to a generalization from inclusion as defined by the binary ordering relation of the poset to a relative degree of inclusion. The degree of implication can be represented as a *real number* [1, 2] denoted $(x \to y)$ defined within a closed interval. Contrast this notation with $x \to y$, which represents the binary ordering relation $x \leq y$, "$x$ is included in $y$". For convenience we choose $(x \to y) \in [0,1]$, where $(x \to y) = 1$ represents the maximal degree of implication with $x \wedge y = x$, which is consistent with $x \to y$, and $(x \to y) = 0$ represents the minimal degree of implication, which is consistent with $x \wedge y = O$. Intermediate values of degree of implication arise from cases where $x \wedge y = z$ with $z \neq x$, $z \neq y$ and $z \neq O$. Thus relative degree of implication is a measure relating arbitrary pairs of assertions in the lattice. Since the binary ordering relation of the poset is all that is needed to define the lattice, there does not exist sufficient structure in the lattice framework to define such a measure. Thus we should expect some form of indeterminacy that will require us to impose additional structure on the space. This manifests itself in the fact that the prior probabilities must be externally defined.

Cox derived relations that the relative degree of implication should follow in order to be consistent with the rules of Boolean logic, i.e. the structure of the Boolean lattice. I will briefly mention the origin of these relations; the original work can be found in [1, 2, 13]. From the associativity of the conjunction of assertions, $(a \to (b \wedge c) \wedge d) = (a \to b \wedge (c \wedge d))$, Cox derived a functional equation, which has as a general solution

$$(a \to b \wedge c)^r = (a \to b)^r (a \wedge b \to c)^r, \qquad (1)$$

where $r$ is an arbitrary constant. The special relationship between an assertion and its complement results in a relationship between the degree to which a premise $a$ implies $b$ and the degree to which $a$ implies $\sim b$

$$(a \to b)^r + (a \to \sim b)^r = C, \qquad (2)$$

where $r$ is the same arbitrary constant in (1) and $C$ as another arbitrary constant. Setting $r = C = 1$ and changing notation so that $p(b | a) \equiv (a \to b)$ one sees that (1) and (2) are analogous to the familiar product and sum rules of probability.

$$p(b \wedge c | a) = p(b | a) \, p(c | a \wedge b) \qquad (3)$$

$$p(b | a) + p(\sim b | a) = 1 \qquad (4)$$

Furthermore, commutativity of the conjunction with (3) leads to Bayes' Theorem

$$p(b | a \wedge c) = p(b | a) \frac{p(c | a \wedge b)}{p(c | a)} \qquad (5)$$

These three equations (3)-(5) form the foundation of inductive inference.

# THE LOGIC OF QUESTIONS

*"It is not the answer that enlightens, but the question."*
-Eugene Ionesco (1912-1994)

*"To be, or not to be: that is the question."*
-William Shakespeare, Hamlet, Act 3 scene 1, (1579)

## Defining a Question

Richard Cox [13] defines *a system of assertions* as a set of assertions, which includes every assertion implying any assertion of the set. The *irreducible set* is a subset of the system, which contains every assertion that implies no assertion other than itself. Finally, a *defining set* of a system is a subset of the system, which includes the irreducible set. As an example, consider the lattice $2^3$ in Figure 3 right. To generate a system of assertions, we will start with the set $\{a,b\}$. The system must also contain all the assertions in the lattice which imply both assertion *a* and assertion *b*. These are all the assertions that can be reached by climbing down the lattice from these two elements. In this case, the lattice is rather small and the only assertion that implies the assertions in this set is *O*, the absurdity. Thus $\{a,b,O\}$ is a system of assertions. The irreducible set is simply the set $\{a,b\}$. Last, there are two defining sets for this system: $\{a,b,O\}$ and $\{a,b\}$. Note that in general there are many defining sets. Given a defining set, one can reduce it to the irreducible set by removing assertions that are implied by another assertion in the defining set, or expand it by including implicants of assertions in the defining set, to the point of including the entire system.

Cox defines a question as the system of assertions that answer that question. Why the system of assertions? The reason is that any assertion that implies another assertion that answers a question is itself an answer to the same question. Thus the system of assertions represents an exhaustive set of possible answers to a given question. Two questions are then equivalent if they are answered by the same system of assertions. This can be easily demonstrated with the questions *"Is it raining?"* and *"Is it not raining?"* Both questions are answered by the statements *"It is raining!"* and *"It is not raining!"*, and thus they are equivalent in the sense that they ask the same thing. Furthermore, one can now impose an ordering relation on questions, as some questions may include other questions in the sense that one system of assertions contains another system of assertions as a subset.

Consider the following question: $T$ = *"Who stole the tarts made by the Queen of Hearts all on a summer day?"* This question can be written as a set of all possible statements that answer it. Here I contrive a simple defining set for *T*, which I claim is an exhaustive, irreducible set

$T \equiv \{a = \text{"}Alice\ stole\ the\ tarts!\text{"},\ k = \text{"}The\ Knave\ of\ Hearts\ stole\ the\ tarts!\text{"},$
$\quad m = \text{"}The\ Mad\ Hatter\ stole\ the\ tarts!\text{"},\ w = \text{"}The\ White\ Rabbit\ stole\ the\ tarts!\text{"}\}.$

This is a fun example as it is not clear from the story[1] that the tarts were even stolen. In the event that no one stole the tarts, the question is answered by no true statement and is called a *vain question* [13]. If there exists a true statement that answers the question, that question is called a *real question*. For the sake of this example, we assume that the question *T* is real, and consider an alternate question *A* = "*Did or did not Alice steal the tarts?*" A defining set for this question is

$$A \equiv \{\, a = "Alice\ stole\ the\ tarts!",\ \sim a = "Alice\ did\ not\ steal\ the\ tarts!" \,\}.$$

As the defining set of *T* is exhaustive, the statement $\sim a$ above, which is the complement of $a$, is equivalent to the disjunction of all the statements in the irreducible set of *T* except for $a$, that is $\sim a = k \vee m \vee w$. As the question *A* is a system of assertions, which includes all the assertions that imply any assertion in its defining set, the system of assertions A must also contain *k*, *m* and *w* as each implies $\sim a$. Thus system of assertions *T* is a subset of the system of assertions *A*, and so by answering *T*, one will have answered *A*. Of course, the converse is not generally true. In the past has been said [11] that the question *A includes* the question *T*, but it may be more obvious to see that the question *T answers* the question *A*. As I will demonstrate, identifying the conjunction of questions with the meet and the disjunction of questions with the join is consistent with the ordering relation "*is a subset of*". This however is dual to the ordering relation intuitively adopted by Cox, "*includes as a subset*", which alone is the source of the interchange between conjunction and disjunction in identifying relations among assertions with relations among questions in Cox's formalism.

With the ordering relation "*is a subset of*" the meet or conjunction of two questions, called the *joint question*, can be shown to be the intersection of the sets of assertions answering each question.

$$A \wedge B \equiv A \cap B. \qquad (6)$$

It should be noted that Cox's treatment dealt with the case where there the system was not built on an exhaustive set of mutually exclusive atomic assertions. This leads to a more general definition of the joint question [13], which reduces to set intersection in the case of an exhaustive set of mutually exclusive atomic assertions. Similarly, the join or disjunction of two questions, called the *common question*, is defined as the question that the two questions ask in common. It can be shown to be the union of the sets of assertions answering each question

$$A \vee B \equiv A \cup B. \qquad (7)$$

According to the definitions laid out in the section on posets, the consistency relation states that *B* includes *A*, written $A \leq B$ (or $A \rightarrow B$) if $A \wedge B = A$ and $A \vee B = B$. This is entirely consistent where the ordering relation is "*is a subset of*", and is dual to the convention chosen by Cox[2] where $B \rightarrow^{\partial} A$ is equated with $A \leq B$ and thus consistent with $A \wedge B = A$ and $A \vee B = B$. As the relation "*is a subset of*" is more conventional, I will deviate here from Cox's convention and say that "*answering A answers B*" or "*B includes A*", written $A \leq B$, or $A \rightarrow B$, when $A \wedge B = A$ and $A \vee B = B$. Although the way in which this relation is expressed is contrary to the handful of published works on inductive logic I make this suggestion to assure that this burgeoning field of inductive logic is notationally and conceptually consistent with the more mature field of lattice theory on which it is undoubtedly based.

---
[1] Chapters XI and XII of <u>Alice's Adventures in Wonderland</u>, Lewis Carroll, 1865.
[2] Highlighting the arrow with a $\partial$ indicates that it is the dual relation, which will be read conveniently as "*B includes A*".

Notation aside, the concepts I have been discussing are unaltered and can be more easily visualized by considering questions *A* and *T* above. The questions "*Who stole the tarts made by the Queen of Hearts all on a summer day?*" and "*Did or did not Alice steal the tarts?*" jointly ask "*Who stole the tarts made by the Queen of Hearts all on a summer day?*" Whereas they ask, "*Did or did not Alice steal the tarts?*" in common. Therefore $T \subseteq A$, which is $T \leq A$, written also as $T \rightarrow A$, read either as "*T answers A*" or "*A includes T*". Dually, *A* includes *T* as a subset, written $A \supseteq T$, which is $A \leq^{\partial} T$, written also as $A \rightarrow^{\partial} T$, and read "*A includes T*".

Next I construct the lattice of questions.

## Ideals and Ideal Questions

An *ideal* is a nonvoid subset *J* of a lattice *A* with the properties [14]

I1.  $a \in J, \; x \in A \;$ where $\; x \leq a \;$ then $\; x \in J$
I2.  $a \in J, \; b \in J \;$ then $\; a \vee b \in J$

In the case that the lattice A is a lattice of assertions, property I1 above is a necessary and sufficient condition for the set *J* to be a system of assertions. Thus each ideal of a lattice of assertions represents a unique system of assertions, or equivalently a question. For this reason, I call these systems of assertions, which are also ideals, *ideal systems* or *ideal questions*.

Given any assertion *x* in the lattice *A*, one can construct the set q(*x*) of all assertions *y* such that $x \leq y$. Thus the function q($\bullet$) takes an assertion to a question. Furthermore, one can show ([14], Theorem 3.3) that the set of all ideals of any lattice *L* ordered by set inclusion forms a lattice $\hat{L}$, and that for a finite lattice $\hat{L}$ is isomorphic to *L*. This is significant, as the space of ideal questions possesses a structure isomorphic to the space of assertions (Figure 5). An inverse mapping can be defined as a function a($\bullet$) that takes an ideal question to an assertion by selecting the greatest element from its system of assertions, so that $a(q(x)) = x$. By virtue of this isomorphism, we know that any identities that hold for the lattice *A* shall also hold for the lattice $\hat{Q}$.

At this point the space of assertions looks isomorphic to the space of questions. However, recall that the ideal questions satisfy an additional property I2, which requires that there be a single greatest element in the set. This is not a property required of questions in general by the definition put forward by Cox. Thus there exist additional questions not represented in the lattice $\hat{Q}$. One such question is the binary question represented by the defining set $\{a, \sim a\}$. If the space of assertions is again $A = 2^3$ then $\sim a = b \vee c$ and the defining set is equivalently $\{a, b \vee c\}$. However, by property I2, the ideal containing the elements in the defining set must also include $a \vee (b \vee c) = a \vee b \vee c$, which is not contained in the system of assertions. Thus the system $\{a, \sim a\}$ is not an ideal question and is not represented in the lattice $\hat{Q}$.

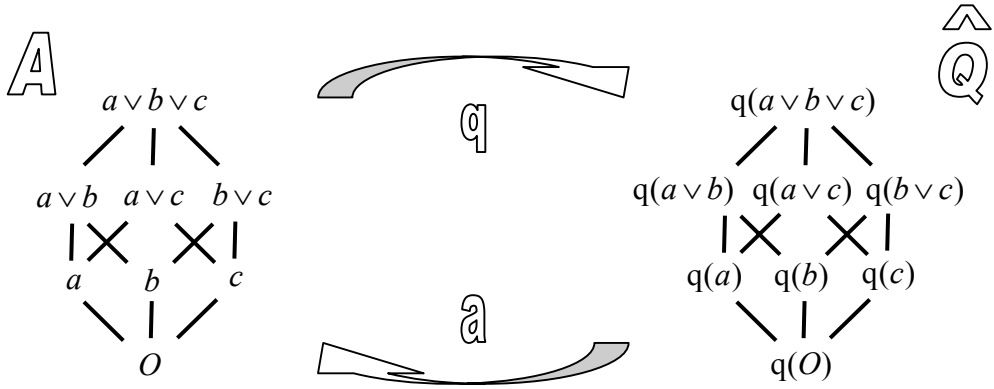

**FIGURE 5.** The lattice of assertions $A \sim 2^3$ (left) and the lattice $\hat{Q}$ (right) obtained by mapping each element $x$ of $A$ to the set q($x$) of all assertions $y \geq x$ and ordering by set inclusion. Note that $\hat{Q}$ and $A$ are isomorphic, written as $\hat{Q} = A = 2^3$.

I now examine the full space of questions in greater detail. As the assertion lattices are $2^N$, I shall also denote the question lattices according to the cardinality of the atomic assertions $N$ by $\mathbf{Q}(N)$, and the lattice of ideal questions is denoted $\hat{\mathbf{Q}}(N) = 2^N$. If a system of assertions defining a question contains an assertion $a$, then the system must contain all the elements of the ideal of $a$, which we have denoted q($a$). Thus any question in the lattice $\mathbf{Q}(N)$ can be constructed from a finite set union of ideal questions from the lattice $\hat{\mathbf{Q}}(N)$. This finite set union can be constructed by using a vector of Boolean values denoting whether or not each of the $2^N$ ideal questions is included in a particular union. The resulting lattice $\mathbf{Q}^N$ is thus the power set $\mathbf{Q}(N) = \wp(\hat{\mathbf{Q}}(N)) = 2^{2^N}$ of $\hat{\mathbf{Q}}(N)$, which is known as the *free distributive lattice* $\mathbf{FD}(N)$ [14, 16]. The lattices $\mathbf{Q}(1)$, $\mathbf{Q}(2)$, and $\mathbf{Q}(3)$ are shown in Figure 6, with notation where $A \equiv q(a)$, $AB \equiv q(a \vee b)$, $ABC \equiv q(a \vee b \vee c)$, and $A \cup BC$ is the set union of the sets $A$ and $BC$. Recall that the natural ordering relation $\subseteq$ of the sets is used.

The number of possible questions grows rapidly with the number of atomic assertions for $N = 1$ through 8: 2, 5, 19, 167, 7 580, 7 828 353, 2 414 682 040 997, 56 130 437 228 687 557 907 787 [16, 17]. The numbers are known as Dedekind's numbers and their determination is known as Dedekind's problem [18]. This is related to the number of monotonic increasing Boolean functions of $N$ variables and to the number of antichains (also called Sperner systems) on the $N$-set [19]. The lattice

**Q**(3) = **FD**(3) with *I* added, (Figure 6, right) is better visualized in three dimensions, and is nicely displayed as an example (FD3) in Ralph Freese's java-based Lattice Drawing Program [20].

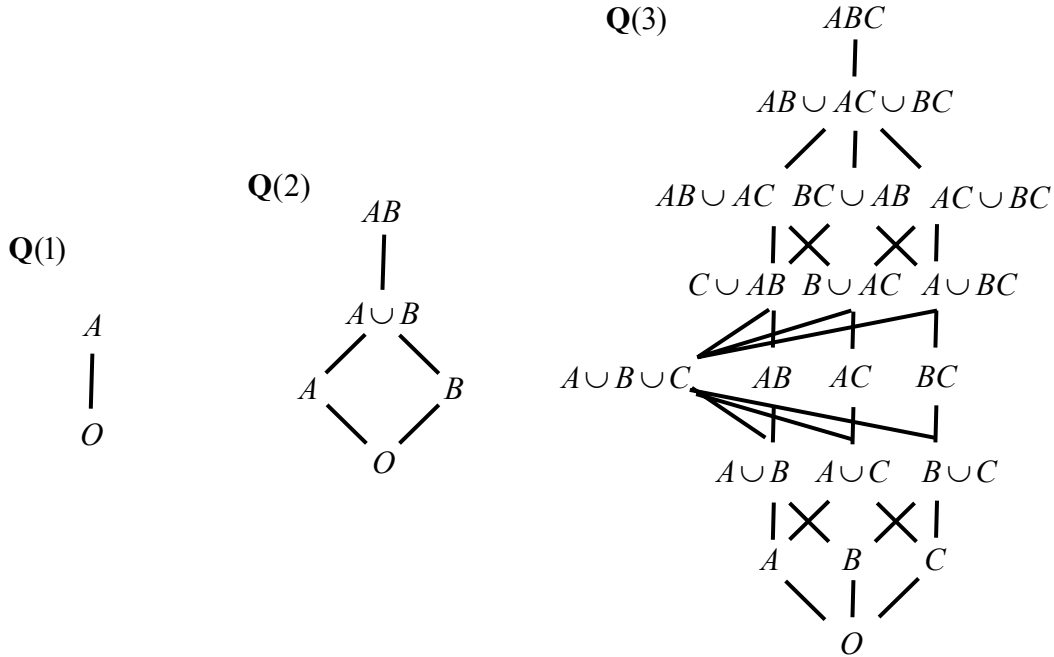

**FIGURE 6**. The question lattices **Q**(1) (left), **Q**(2) (center), and **Q**(3) (right). These lattices are the free distributive lattices with 1, 2, and 3 generators respectively. Note that $A \equiv q(a)$, $AB \equiv q(a \vee b)$, $ABC \equiv q(a \vee b \vee c)$, and $A \cup BC$ is the set union of the system of assertions for questions $A$ and $BC$.

## Real Questions

Thus far in these examinations one important point has been neglected; I have not stipulated that the assertions defining a question be exhaustive. That is, there is no assurance that all of the questions in the lattice **Q**(*N*) are *real questions* answerable by a true assertion. As the atoms of the lattice of assertions $2^N$ are an exhaustive set, then only questions containing the set of atoms as a subset are assured to be real questions. There of course may be questions that do not contain this entire set, that for a given situation may be answerable by a true assertion, but this in general is not guaranteed *a priori*. The least element that contains the set of atoms as a subset is given by $R_\perp = \vee_{i=1}^N q(a_i)$, where $\vee_{i=1}^N q(a_i) = q(a_1) \vee q(a_2) \vee \ldots \vee q(a_N)$, which is the disjunction of all the ideals formed from the *N* atomic assertions. This is $A$, $A \cup B$, $A \cup B \cup C$ for lattices **Q**(1), **Q**(2), and **Q**(3) respectively. Thus all the lattice elements that are greater than this question $R_\perp$ are all assured to be real questions that can be answered by every atomic assertion in the exhaustive set. These *assuredly real questions* are bounded above by the question at the top of the lattice, *I*, which I will instead denote as $R_T = q(\bigvee_{i=1}^{N} a_i)$, where $\bigvee_{i=1}^{N} a_i = a_1 \vee a_2 \vee \ldots \vee a_N$. It can be easily shown that these assuredly real questions bounded by $R_\perp$ and $R_T$ form a sublattice **R**(*N*) where all joins and meets of elements of **R**(*N*) are also elements of **R**(*N*).

Looking at the lattices in Figure 6, it appears in each case that the sublattice **R**(*N*) (excluding $R_T$) is Boolean (compare to the lattice structures in Figure 3). However, this pattern does not hold in general and in fact fails for **Q**(4). This can be demonstrated by looking at what are called the join-irreducible elements of **R**(*N*). In short these are the elements of a lattice that cannot be written as a join of elements of the lattice, excluding *O*. In any finite Boolean lattice, the join-irreducible elements are its atoms (see $2^3$ in Figure 3) [21, 22]. The poset formed by these atoms alone consists only of these elements side-by-side, and is called an antichain (Figure 7a). Thus the join-irreducible elements of a Boolean lattice form an antichain, written symbolically as $J(2^N) = \overline{\mathbf{N}}$. A proof that **R**(*N*) is not Boolean, which will be published by in a future paper, relies on the observation that the join-irreducible elements of **R**(*N*) are of the form $\bigvee_{i=1}^{M} q(a_{b_i}) \vee q(\bigvee_{j=M+1}^{N} a_{b_j})$ where $a_k$ represents the $k^{\text{th}}$ atom of $2^N$ from which **R**(*N*) is formed and *b* is some permuted sequence of the set of natural numbers from 1 to *N*, and $1 \leq M < N$. In **R**(3) there are three join-irreducible elements $\{A \cup BC, B \cup AC, C \cup AB\}$, which form an antichain and hence **R**(3) (excluding $R_T$) is a Boolean lattice. In **R**(4) there are a total of 10 join-irreducible elements: 4 of $\{A \cup BCD, \cdots, D \cup ABC\}$ and 6 of $\{A \cup B \cup CD, A \cup C \cup BD, \cdots, C \cup D \cup AB\}$. However, these 10 elements do not form an antichain since $A \cup B \cup CD \leq A \cup BCD$, and so on. Figure 7 shows the forms of $J(\mathbf{R}(3))$, $J(\mathbf{R}(4))$ and $J(\mathbf{R}(5))$. The fact that **R**(*N*) is not in general a Boolean lattice has a very important implication – its elements are not complemented. Therefore, assuredly real questions, like questions in general, do not possess complements.

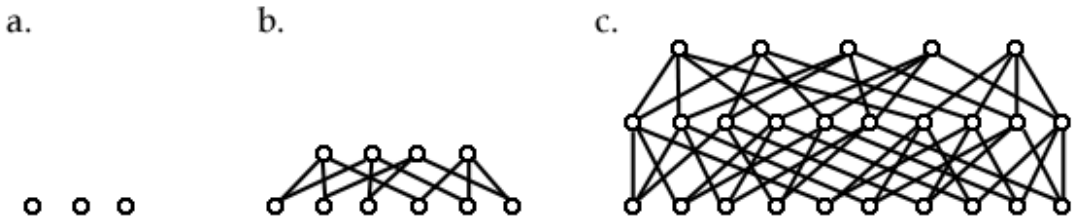

**FIGURE 7.** The join-irreducible elements of the sublattice of real questions (excluding $R_T$), (a.) $J(\mathbf{R}(3)) = \overline{\mathbf{3}}$ is an antichain, thus **R**(3) excluding $R_T$ is a Boolean lattice, whereas (b.) $J(\mathbf{R}(4))$, and (c.) $J(\mathbf{R}(5))$ are not antichains indicating that **R**(4), **R**(5) and in general **R**(*N*) are not Boolean lattices. Drawing these structures in a tidy way is quite a challenge. Note that I have not labeled the elements (described in the text for **R**(3) and **R**(4)) and that their ordering in the diagram is not necessarily the order of the listing the text.

# Inductive Inquiry on Lattices

As briefly described earlier, the sum rule of probability (2) derives from the fact that Boolean lattices are uniquely complemented. In Cox's earlier work, he described how there could be no complete analog in the algebra of systems (questions) to the complement in the algebra of assertions ([2], pp. 52-3). In a footnote Cox describes how Boole [23] applied his algebra to classes of objects in addition to propositions (see Figure 1). He notes that one might be inclined to think of a system as a class as defined by Boole, however the set of assertions not included in a system, while forming a class, do not itself form a system. For this reason the algebra of systems cannot possibly be Boolean.

In Cox' paper on inquiry [13] he defines a mutually contradictory pair of questions "as a pair whose conjunction is equal to the conjunction of all questions, and whose disjunction is equal to the disjunction of all questions." While this definition is acceptable, he does not prove their existence. While my discussion on join-irreducible elements may convincingly prove to a mathematician familiar with the theory that complements to questions do not exist, those less-familiar may require more tangible evidence. Consider the question $A \cup B$ in the lattice $\mathbf{Q}(3)$, (Figure 6c). Its hypothetical complement must satisfy two relations $(A \cup B) \vee \sim(A \cup B) = I$ and $(A \cup B) \wedge \sim(A \cup B) = O$. Consider the first relation $(A \cup B) \vee \sim(A \cup B) = I$. If its complement $\sim(A \cup B) > (A \cup B \cup C)$ then $\sim(A \cup B) > (A \cup B \cup C) > (A \cup B)$, which by the consistency relation gives $\sim(A \cup B) \vee (A \cup B) = \sim(A \cup B)$. This implies that its complement is $I$, which is a contradiction. Now $(A \cup B \cup C)$ cannot be its complement as $(A \cup B \cup C) > (A \cup B)$. So its complement must satisfy $(A \cup B \cup C) > \sim(A \cup B)$. However, $\sim(A \cup B) \vee (A \cup B) \leq (A \cup B \cup C)$ as both $(A \cup B \cup C) > (A \cup B)$ and $(A \cup B \cup C) > \sim(A \cup B)$, which is again a contradiction. Thus there does not exist a complement to the question $A \cup B$ in the lattice $\mathbf{Q}(3)$.

Last, distributive lattices share the associative and commutative properties of the Boolean lattice. For this reason, one can fully expect that generalizations of the binary ordering relation to measures of degree of inclusion will result in a calculus possessing a product rule as well as a rule analogous to Bayes' Theorem.

## RELEVANCE AND PROBABILITY

There is a deep relationship between the Boolean lattice and the free distributive lattice generated from it. Looking at the lattices $\mathbf{Q}(1)$, $\mathbf{Q}(2)$, and $\mathbf{Q}(3)$, one can see that the join-irreducible elements are precisely the ideal questions, which have a lattice structure isomorphic to the original Boolean lattice from which the questions were generated. This is the map[3] $Q \mapsto J(Q)$, whereas the process of generating the question lattice is a map from the Boolean lattice of assertions to the question lattice, which I write as $A \mapsto O(A)$. We thus have an isomorphic correspondence between the lattice structures where $Q = O(A)$ and $A = J(Q)$. This is true in general for all finite distributive lattices $Q$ and all finite ordered sets $A$ and is known as Birkhoff's Representation Theorem [16]. The lattice $Q$ is called the *dual* of $J(A)$ and $A$ is called the *dual* of $O(Q)$, however this duality should not be confused with the duality induced by the ordering relation discussed earlier. Furthermore, it can be shown that the join-irreducible map takes products of lattices to sums of lattices, so that one can think of $Q \mapsto J(Q)$ and $A \mapsto O(A)$ as being the logarithm and exponential functions, respectively, for lattices [16]. This is quite enticing in that it further supports our expectations that the relevance of a question on an issue can be represented in terms of the logarithms of the probabilities of the assertions involved, and that entropy may play the same role with distributive lattices as probability does with Boolean lattices.

---

[3] Note that these are maps from one lattice structure to another, and are not maps from an element in one lattice to an element in another.

## THE ROLE OF ORDER

The lattice structure of the space of assertions and the space of questions has provided great insights into their structures, symmetries, and relationships. In addition, the associative and commutative properties of lattices suggest that analogies to the familiar product rule of probability and Bayes' Theorem may appear in the calculi of other fields where ordering relations play an important role. This in fact may have already been recognized with the realization that the cross-ratio in projective geometry has the same form as the odds ratio from Bayes' Theorem [24]. Considering the findings in this paper, such a relationship may no longer be such a mystery as the notion of closeness in a projective space provides such an ordering relation. In fact, we might now not be surprised to see forms identical to probability and perhaps entropy appearing in seemingly unrelated fields. In such cases, it is not geometry that underlies these theories – but order.

"*The important thing is not to stop questioning.*"
-Albert Einstein (1879-1955)

## ACKNOWLEDGEMENTS

I would like to thank Jeffrey Jewell for noting at MaxEnt 2001 that my work reminded him of lattice theory, and Ralph Freese, David Clark, and Mick Adams for their correspondences, which have helped me to better understand the ins-and-outs of this theoretical framework. I would also like to extend a special thank you to my friend and colleague Domhnull Granquist-Fraser who inspired me to keep my focus during this effort. Last, I would like to thank Bob Fry for introducing me to this fascinating area of study and for his continued friendship.